\documentclass[preprint,authoryear,1p]{elsarticle}

\usepackage{amssymb}
\usepackage{graphicx}
\journal{Icarus}

\begin{document}
\begin{frontmatter}

\title {Physical studies of Centaurs and Trans-Neptunian Objects  with the Atacama Large Millimeter Array}

\author[cfa]{Arielle Moullet}, 
\author[meu]{Emmanuel Lellouch}, 
\author[meu]{Raphael Moreno}, 
\author[cfa]{Mark Gurwell}

\address[cfa]{Harvard-Smithsonian Center for Astrophysics, Cambridge MA-02138 (U.S.A.)}
\address[meu]{LESIA-Observatoire de Meudon}

%% ----- END ELSEVIER STUFF -----

%\begin{flushleft}
%\vspace{1cm}
%Number of pages: \pageref{lastpage} \\
%Number of figures: \ref{lastfig}\\
%\end{flushleft}

%\begin{pagetwo}{Physical studies of Centaurs and TNOs with ALMA}%                        
%1         2         3         4         5
%               1234567890123456789012345678901234567890123456789012345

%Arielle Moullet\\
%Harvard-Smithsonian Center for Astrophysics\\
%MS 78\\
%Cambridge, MA-02138, USA. \\
%\\
%Email: amoullet@cfa.harvard.edu\\
%Phone: (617) 495-7085 \\
%Fax: (617) 495-708

%\end{pagetwo}

\begin{abstract}
Once completed, the Atacama Large Millimeter Array (ALMA) will be the most powerful (sub)millimeter interferometer in terms of sensitivity, spatial resolution and imaging. This paper presents the capabilities of ALMA applied to the observation of Centaurs and Trans-Neptunian Objects, and their possible output in terms of physical properties. Realistic simulations were performed to explore the performances of the different frequency bands and array configurations, and several projects are detailed along with their feasibility, their limitations and their possible targets. Determination of diameters and albedos via the radiometric method appears to be possible on $\sim$500 objects, while sampling of the thermal lightcurve to derive the bodies' ellipticity could be performed at least 30 bodies that display a significant optical lightcurve.
On a limited number of objects, the spatial resolution allows for direct measurement of the size or even surface mapping with a resolution down to 13 milliarcseconds.
Finally, ALMA could separate members of multiple systems with a separation power comparable to that of the HST.
The overall performance of ALMA  will make it an invaluable instrument to explore the outer solar system, complementary to space-based telescopes and spacecrafts.
\end{abstract}

% %% Keywords should appear after the abstract. 
\end{frontmatter}
%% ----- END ICARUS STUFF -----

%main text
\section{Introduction}
Almost 1400 small bodies orbiting beyond Jupiter have been discovered so far, classified as 
Centaurs (orbiting within Neptune's orbit) or as Trans-Neptunian Objects (TNOs, orbiting beyond Neptune). Due to their distance to the Sun and hence their low physical and chemical processing \citep{mckinnon2008}, their surfaces are expected to expose some of the most pristine material in the solar system. To relate today's outer solar system characteristics to those of the primordial disk, we need to better constrain physical properties and composition of these bodies, and understand the physical, chemical and dynamical processes that took place to shape this region.\\
 Obtaining thermal emission measurements on these bodies is an essential tool for this purpose, since they can 
give access to properties such as geometric albedo, size, shape and surface properties (e.g.
thermal inertia, emissivity). Building a large database of albedos and sizes is necessary to identify possible 
correlations with spectral properties (infrared and visible colors) or dynamical parameters, that would help to understand the roles of processes such as space weathering and collisions \citep{doressoundiram2008}. This would also allow refinement of the taxonomy of this population, and to compare it to other populations (comets, asteroids) so as to identify similarities and to trace population histories. Constraints on the size distribution power law can give clues on the planetesimal growth and fragmentation processes \citep{kenyon1999}, 
and the identification of breaks in the size distribution is a powerful diagnostic of the intrinsic strength of these bodies \citep{pan2005}.  Accurate determination of the shape and size of individual bodies,
along with mass determination, is important to determine their formation and collisional history, 
as well as their bulk density and their ability to retain an atmosphere and/or surface ices \citep{lacerda2007,levi2009,schaller2007}. Knowledge of surface albedos is also necessary to correctly interpret the ice bands that can be detected in near-infrared and visible spectra, and thus to accurately establish surface composition \citep{barucci2008}. Measuring the variation of spectral emissivity with wavelength, by combining {\em Herschel} and ALMA data, gives access to thermophysical and composition properties of the surface and subsurface. Finally, precise determination of the brightness temperature itself is a key indicator of the physical temperature of the surface and subsurface, establishing the possible presence and abundance of a stable atmosphere sustained by ice sublimation.
\\
So far, only about 50 Centaurs and TNOs with diameters generally larger than 100~km have been detected at thermal wavelengths, mostly by space-based infrared 
telescopes ISO \citep{thomas2000} and especially {\em{Spitzer}} \citep{stansberry2008,brucker2009}. With disk-averaged surface temperatures below 130~K for Centaurs and 50~K for TNOs, their thermal emission peaks between 20-100 microns, where the Earth's
 atmosphere is opaque. At longer wavelengths, the thermal emission is considerably lower. Only 8 bodies have been detected in the (sub)mm wavelength range \citep{altenhoff1995,altenhoff2001,margot2002,lellouch2002,altenhoff2004,bertoldi2006,gurwell2010}, mostly around 250~GHz (1.2~mm) with the MAMBO bolometer on the IRAM-30m antenna. Interferometric facilities, with smaller or comparable total collective area, could only detect the brightest bodies (Pluto and Charon) so far \citep{gurwell2010}, but the recent correlator upgrades on the IRAM-Plateau de Bure array now allows for point-source sensitivities better than MAMBO at 250~GHz. \\
The number of thermal detections is expected to increase significantly in the coming years. At 70-160 microns, {\em Herschel}'s sensitivity may allow to detect up to 140 Centaurs and TNOs, that are the targets of 
a large photometric program \citep{mueller2009}; first 
results have just been presented \citep{mueller2010,lim2010,lellouch2010}.
In the (sub)mm wavelength range, starting in 2012-2013, the Atacama Large Millimeter Array (ALMA) will 
provide unprecedented sensitivity, allowing in principle the detection of a dramatically larger number of targets. This interferometric facility, under construction in Chile, will offer at completion 50 antennas of 12~m diameter each, along with 12 additional 8-m antennas and four 12-m antennas forming the Atacama Compact Array (ACA). In addition, the array will offer very extended configurations 
with baselines up to 14 km, that will provide spatial resolution down to 5 milliarcseconds (mas) at 850~GHz (350~microns), corresponding to $\sim$150~km at 40~AU, 
which as we will show is sufficient to resolve the largest Centaurs and TNOs.  \\
\\
This paper presents an analysis of the capabilities of ALMA for physical studies of Centaurs and TNOs, in terms of sensitivity, resolution and imaging performance. Detailed simulations were performed, which take into account the array characteristics, atmospheric quality, expected receiver performance and correlator capabilities. 
Imaging capabilities were simulated using a realistic simulator, developed by the GILDAS team \citep{pety2002}, to calculate the expected Fourier-plane coverage. Feasibility and observing strategies for a number of detection and imaging projects are detailed, along with their expected products
in terms of the bodies' properties and in general outer solar system science. A short study on ALMA capabilities for asteroid science can be found in \citet{busch2009}.

\section{ALMA technical characteristics and expected performance}
\subsection{Point-source sensitivity}
To characterize the expected noise for a given observation, the point source sensitivity is commonly used in radio astronomy. In the case of continuum emission measurements, this corresponds to the rms noise expected on flux measurements on an unresolved source obtained using the whole bandwidth of the instrument $\Delta \nu$ in $\Delta t$ seconds of time. This quantity depends on the instrumental performances (antennas, receivers, correlator) coupled with the atmospheric qualities of the site (sky opacity and phase stability), following the classical formula expressed in flux density units \citep{thompson}:\\
\begin{equation}
\Delta (S)(Jy)= \frac{K T_{sys}}{\eta_{atm}\eta_{cor} \sqrt{\Delta t \Delta \nu n_p N (N-1)}}
\end{equation}
where  $\eta_{cor}$ is the correlator efficiency, $T_{sys}$ the system temperature characterizing the receiver and sky noise, $n_p$ the number of polarizations and N the number of antennas. The $K$ term describes the gain of the antennas in Jy/K, and is defined as $\frac{2 k F_{eff}}{A_{col} \eta_{a}}$ with $k$ the Boltzmann constant, $F_{eff}$ the forward efficiency, $A_{col}$ the collecting area of a single antenna, and $\eta_{a}$ the aperture efficiency (e.g. $K=40$~Jy/K at 230~GHz at the IRAM-PdBI). Finally, $\eta_{atm}$ is the phase decorrelation (equal to e$^{-\sigma^2/2}$, $\sigma$ being the phase rms at the observing frequency), that measures the effective signal loss due to the atmospheric phase fluctuation. \\
To calculate the point-source sensitivity expected from the ALMA array for each observing frequency, we will use the following approximate equation from \citet{debreuck2005} for dual polarization observations ($n_p$=2),
where the correlator and forward efficiencies are considered independent from frequency :
\begin{equation}
\Delta (S)(mJy)=\frac{2.6\times10^6 T_{sys}}{\eta_{a} \eta_{atm}N D^2 \sqrt{\Delta t \Delta \nu}}
\end{equation}
We will consider the other parameters as following :\\
- N=50, D=12~m : we thus consider only the main array at completion, excluding the contribution of the adjacent ACA (Atacama Compact Array). \\
- $\Delta \nu$=8~GHz per polarization using the full correlator capacity.\\
- $\eta_{atm}$=0.87, corresponding to a phase rms of 30$^{\circ}$\\
- $\eta_{a}$ : this parameter depends mostly on the accuracy $\sigma_{a}$ of the surface of the antennas. Following \citep{ruze1966}, $\eta_{ant}=\eta_0 e^{-(\frac{4\pi\sigma_{a}}{\lambda})^2}$, where $\lambda$ is the observed wavelength, and $\eta_0$ is the efficiency of a perfectly smooth antenna. The latest measurements on the already available antennas show that $\sigma_{a}$ is of the order of 20~$\mu$m although it could rise 
to 25~$\mu$m in bad conditions (cold weather) (Wooten, private comm). The value of $\eta_0$ is assumed to be 
0.8 following the antenna requirements as defined in \citet{butler99}.\\
- system temperature (T$_{sys}$, in K) : this parameter includes the combined effects of the thermal noise from the receivers and the opacity of the atmosphere. The values assumed here are the estimates by \citet{moreno2002}, calculated for a source at 50$^{\circ}$ elevation, and assuming
a water vapor content varying with the observing frequency (0.5~mm for frequencies above 370~GHz and either 2.3~mm or 1.2~mm below), which is realistic since high-frequency observations require better sky conditions. The T$_{sys}$  were lowered for bands 8 and 9 by respectively 33\% and 40\%, to match the latest sensitivity expectations as specified by the ALMA sensitivity simulator (http://www.eso.org/sci/facilities/alma/observing/tools/etc/). In addition, for band 6, the estimates have been updated using the most recent performance measurements on the receivers (ALMA Newletter, September 2010, www.almaobservatory.org). \\
\\
The values of antenna efficiency, system temperature and point-source sensitivity, derived from Equation 2, are gathered in Table 1 for a set of characteristic frequencies. For comparison, the best continuum point-source sensitivities reached by available instruments are shown. ALMA should provide very significant gains in sensitivity (factors 10-100). However given the minimal characterization of the system so far, the ALMA sensitivity estimates may change as new 
performance measurements will become available. It is also probable that in average, 10\% of the antennas may not be used during standard observations, due to maintainance and/or 
technical failures. This would degrade the sensitivity estimates by 10\%.\\
Finally we note that the point-source sensitivity should not strongly depend on the array configuration. Indeed, although the atmospheric decorrelation typically increases as the distances between antennas increase (i.e. with baseline length), the ALMA antennas will be equipped with  water vapor radiometers at 183~GHz, that will monitor variations of the atmospheric optical depth along the line-of-sight. Application of the derived  phase corrections should help to compensate the atmospheric decorrelation.  
Keeping these limitations in mind, we will assume the continuum point-source sensitivities presented in Table 1 in the rest of the paper.
\begin{table}
\begin{center}
\begin{tabular}{|c|c|c|c|c|c|}
\hline
Frequency  & Wavelength & $\eta_{a}$ & T$_{sys}$ & ALMA Sensitivity & Best present sensitivity \\
(GHz) & (mm) & & (K) & (mJy/hour) & (mJy/hour)\\
\hline
110 (band 3) & 2.7 & 0.79 & 81 & 0.008 & 0.15 (Plateau de Bure-IRAM)\\
230 (band 6) & 1.3 & 0.77 & 118 & 0.012 & 0.31 (Plateau de Bure-IRAM)\\
345 (band 7) & 0.87 & 0.74 & 222 & 0.023 & 2.8 (LABOCA-APEX)\\
460 (band 8) & 0.65 & 0.69 & 460 & 0.051 & \\
675 (band 9) & 0.44 & 0.58 & 636 & 0.085 & 1.6 (SCUBA2-JCMT, expected)\\
850 (band 10) & 0.35 & 0.48 & 1200 & 0.19 & 41 (SABOCA-APEX)\\
\hline
\end{tabular}\vspace{0.5cm}
\caption{Estimated aperture efficiencies, system temperature and continuum point source sensitivities for ALMA, for 1 hour observation on-source with the whole 16~GHz equivalent bandwidth, derived from Equation 2. The best sensitivity achieved by present facilities are indicated for comparison.
\label{Tablesens}}
\end{center}
\end{table}

\subsection{Array configurations and imaging performance}
A significant advantage of interferometers when compared to single-dish instruments is their ability to boost spatial resolution. While single-dishes are diffraction-limited by the size of the antenna, interferometers are diffraction-limited by the length of their longest baseline, i.e. the longest distance between two antennas. The resolution unit of an interferometer is the synthesized beam, which is the Fourier transform of the spatial distribution of the antennas,
and whose size is characterized by the half power beam width (HPBW). The extent of the baselines is however limited by terrain, atmospheric phase decorrelation and lack of phase correction. Baselines up to $\sim$1~km only are offered on the present interferometers.\\
The ALMA interferometer has been designed to offer flexible configurations of its 50 antennas, with baselines 
up to 14~km. This is possible thanks to the good atmospheric stability on the site and the phase
correction allowed by the 183~GHz water radiometers. The expected resolution for different bands as a function of observing frequency is shown in Figure \ref{resolution}, along with the resolution available from other facilities. An angular resolution down to 5~mas is expected for the highest frequencies, and down to 20~mas at 230~GHz, which is better by a factor 14 than what the Combined Array for Research in Millimeter Astronomy (CARMA) can reach.\\
\\
However, high spatial resolution is not sufficient to ensure good quality imaging. Indeed, an interferometer measures samples of the Fourier transform of the sky brightness distribution, the visibilities. Each visibility corresponds to the correlation of the signal coming from a pair 
of antennas at a given time. The coordinates $(u,v)$ of the visibilities in the Fourier-plane (or spatial frequencies plane) depend on the source local coordinates on the sky, the baseline length and its orientation. To retrieve the original brightness distribution, the obtained set of visibilities must undergo an inverse Fourier transform. This process is all the more reliable when the coverage of the Fourier-plane is as complete as possible which implies baselines of different lengths and orientations.\\
Thanks to its high number of baselines, low latitude (-23$^{\circ}$S) and configuration design, the ALMA array will offer, in a very short time, very satisfying Fourier-plane filling, sufficient to ensure a good distribution of visibility data in all directions and spatial frequencies, out to a maximum distance. Figure \ref{uvcov} gives an example of Fourier-plane coverage obtained in just 15 minutes, {and the corresponding synthesized beam. The array thus
provides almost instantaneous good imaging quality, enabling time-resolved imaging on thermally bright and large enough sources.

\begin{figure}
\includegraphics[width=15cm]{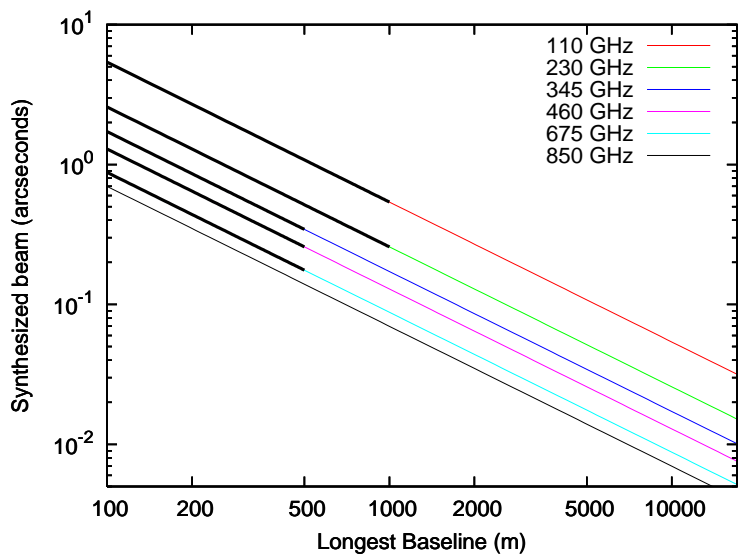}
\caption{\label{resolution} Spatial resolution offered by ALMA,  defined by the synthesized beam half power width, as a function of the longest baseline of the configuration and the operating frequency. Best spatial resolutions offered by other facilities (CARMA at 110 and 230~GHz, SMA at 345, 460 and 675~GHz) are represented with thick black lines.}
\end{figure}

\begin{figure}
\begin{minipage}{18cm}
\includegraphics[width=7cm,angle=-90]{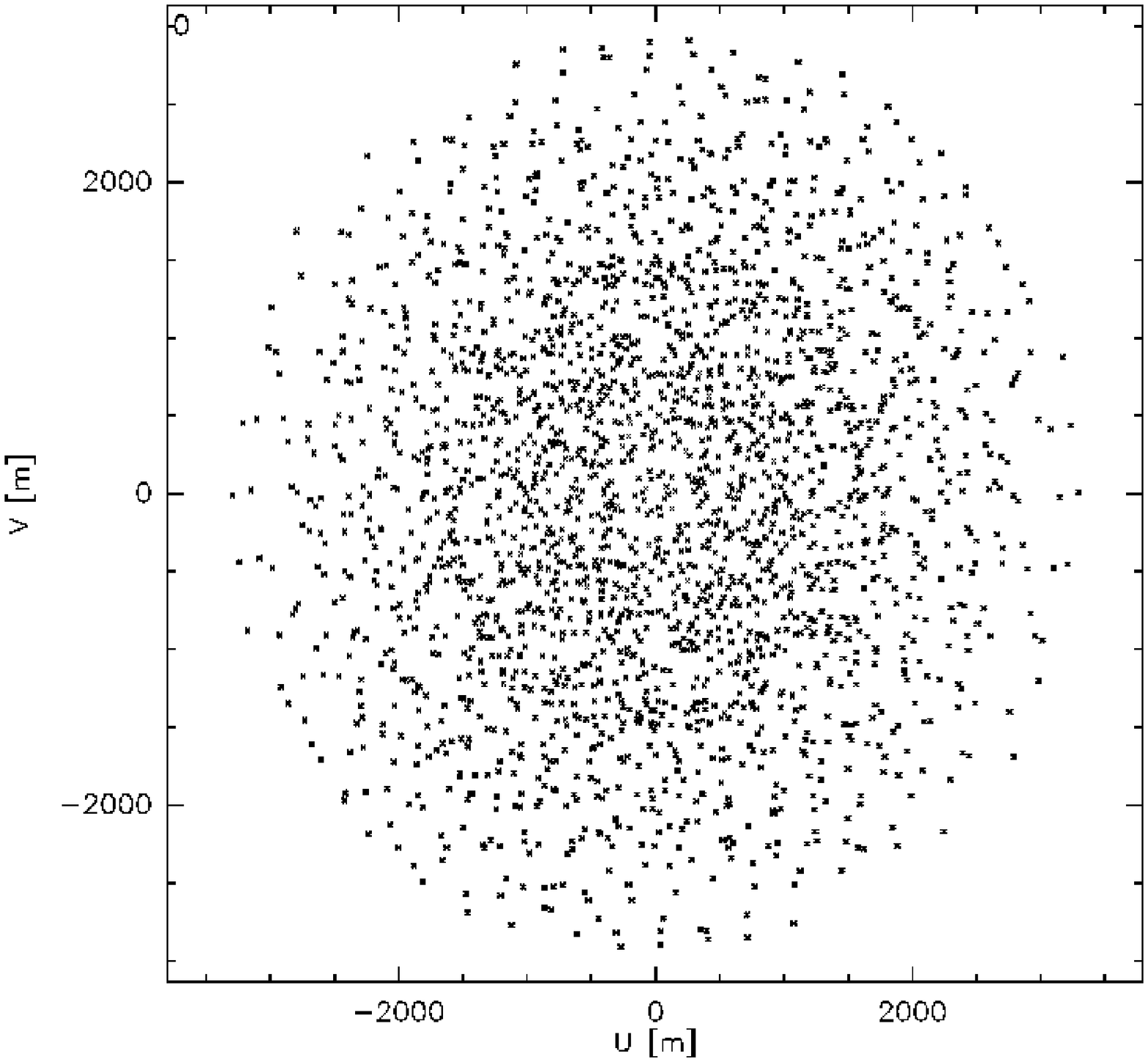}
\includegraphics[width=7cm,angle=-90]{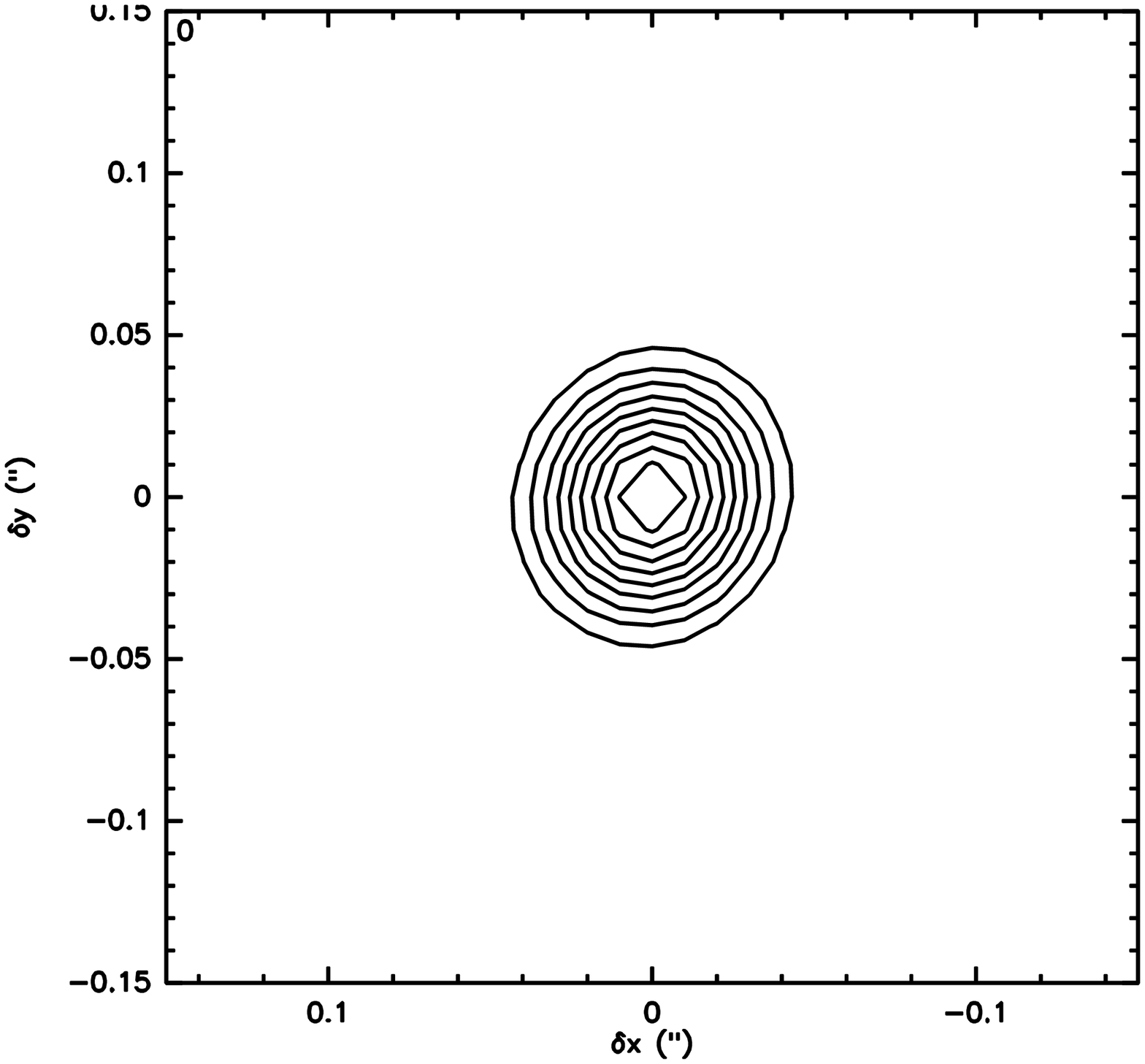}
\end{minipage}
\caption{\label{uvcov} Left : Fourier-plane coverage obtained in 2 minutes on a -10$^{\circ}$ declination transiting source, with a simulated extended configuration of ALMA. Each point corresponds to a visibility, i.e. the cross correlation of the signals from a pair of antennas, whose coordinates in the Fourier plane represent the length and orientation of the corresponding projected baseline, in meters. Right : corresponding synthesized beam, of $\sim$ 50~mas HPBW. Axes are expressed in arcseconds.}
\end{figure}

\section{ALMA performance for Centaurs and TNOs observations}
\subsection{Adopted thermal emission models}
To assess the performance of ALMA for the observation of outer solar system small bodies, a model of their thermal emission is required. However to fully describe it, one needs 
information on many geometric and physical properties (pole orientation, thermal inertia, albedo...) that are 
still unknown in most cases. The best approximate approach is to use empirical thermal models designed to reasonably reproduce the thermal emission from Centaurs and TNOs, and, for some parameters, to assume 
typical values. The models and assumptions used are described below.\\
\\
In the grey-body emission model, for a single spherical body of 
apparent radius $R$ (in radians), the thermal flux density emitted at a frequency $\nu$ from a  body corresponds to the Planck law :
\begin{equation}
F(\nu)=\epsilon \frac{2h\nu^3}{c^2} \pi R^2\int{\frac{sin(\phi) 
d\phi d\theta}{e^{\frac{h\nu}{kT(\phi,\theta)}}-1}}
\label{equflux}
\end{equation}
where $h$ is Planck's constant, $k$ is Boltzmann's constant, $\phi$ and $\theta$ are respectively the 
local latitude and longitude on the body, and  $T(\phi,\theta)$ is the surface temperature distribution.
\\
The spectral emissivity $\epsilon$ characterizes the departure from black-body
emission, which varies with wavelength over the (sub)mm range. In particular, $\epsilon$ takes into account that the thermal emission is not entirely reemitted towards space, depending on the dielectric constant of the surface. It also reflects the fact that a 
surface is not completely opaque at thermal wavelengths (i.e. the absorption coefficient by the surface
is finite), which means that the outgoing emission is the sum of contributions from different depths below the surface. To phenomenologically represent these aspects, we use an emissivity factor varying 
between 0.8 to 1, consistent with the emissivities found by \citet{moreno2009} on Callisto and Ganymede. The emissivity is assumed to be independent of the position on disk, i.e. we do not consider a Fresnel-like description of emissivity but rather a Lambertian surface model.\\
The temperature distribution on a body depends on the local solar illumination and on the surface response to it. It can be derived from the equations of heat conduction if all geometric (solar distance, latitude of sub-earth point, rotation rate) and physical parameters (albedo, bolometric emissivity, thermal inertia) are known. 
This is however generally not the case. Instead we have considered three different empirical thermal models, 
adapted from \citet{spencer89}:\\
- a "hot" model, the Equilibrium Model (EQM). This describes a body with zero thermal inertia  (or a very slow 
rotator). In this model, the temperature varies only with the insolation angle (solar zenith angle, SZA) 
as $T=T_{ss} cos^{0.25}(SZA)$, where $T_{ss}$ is the temperature at the sub solar point, defined as $T_{ss}= (\frac {(1-A_b)F}{r_h^2 \epsilon_{b} \sigma})^{1/4}$, where $F$ is the solar constant at 1~AU (in W.m$^{-2}$), $r_h$ is the heliocentric distance, $A_{b}$ is the bolometric albedo, and $\epsilon_{b}$ is 
the bolometric emissivity, that we will assume equal to 0.9. The bolometric albedo is the product of the 
phase integral $q$, assumed equal to the standard value for asteroids (0.4), with the geometric albedo $p_v$, for which, if unknown, we will assume the average value of 0.08 from \citet{stansberry2008}. This assumption is valid for the majority of intermediate and small-sized bodies, while very large bodies tend to present higher albedos, up to 0.9 (e.g. Eris, \citet{brown2006}).\\
- a "cold" model, the Isothermal Latitude Model (ILM), equivalent to a body with infinite thermal inertia 
(or a rapid rotator). In that case, temperature varies only with latitude, as $T=\frac{T_{ss}}{\pi^{0.25}} 
cos^{0.25}(\phi)$, with $T_{ss}$ as defined above.\\
- an intermediate model, the Standard Thermal Model (STM), whose distribution on the disk follows that 
of the EQM, but with a corrective factor $\eta^{-1/4}$. The $\eta$ parameter  allows in an empirical way to 
situate the thermal model of a given body in between the hot and cold models. Observationally,
$\eta$ can be constrained from measurements at different wavelengths. \\
In this paper, we will use as reference model, the "TNO-tuned" STM model, i.e. with $\eta$ equal to 1.25, 
the average value determined from {\em{Spitzer}} data \citep{stansberry2008}, and a spectral emissivity $\epsilon$=0.9. The other two models will be used to estimate the range of plausible temperature distributions.

\subsection{Detection thresholds}
A number of projects that can be performed with ALMA are based on the measurement of the total thermal flux to derive a number of parameters. Detection of the source at 5~$\sigma$ is the main requirement to perform  those projects, and is also a pre-requisite condition to perform all other projects described in the following sections. We want to establish how many Centaurs and TNOs are potentially detectable by ALMA, assuming that their thermal emission follows the adopted reference thermal model.
We will consider as detectable a body whose thermal emission corresponds to at least 
5 times the rms noise (5~$\sigma$ detection). 
Figure \ref{detection} displays the detection limit at 6 different frequencies, for a 1-hour on-source integration. This integration time is 
reasonable for large surveys, 
but longer integration times are possible for specific sources. Figure \ref{detection} shows that in
the most sensitive band (band 7, around 345~GHz), one could detect bodies with equivalent diameters  
larger than 75~km at 20~AU, 130 km at 30~AU, 190~km at 40~AU and 250~km at 50~AU. Bands 6 and 9 (respectively 230~GHz and 675~GHz)
are slightly less 
efficient than band 7, and other bands are significantly less sensitive : at 40~AU, only bodies larger 
than 250~km can be detected with band 10 (850~GHz). \\
To illustrate this, 1360 known objects are plotted on Figure \ref{detection}. For most of them, the 
radius is unknown, and was estimated from their visual magnitudes (taken from 
http://vo.imcce.fr/webservices/miriade), 
assuming a geometric albedo of 0.08. 
Figure \ref{detection} shows that more than 500 Centaurs and TNOs could be detected with band 7, 460 with band 6, 430 with band 9 and 330 with band 8 (460~GHz) in 1-hour on-source observation per object. \\
\\
The sensitivity performance of ALMA clearly increases the potential number of detectable targets, considering that only $\sim$50 bodies were detected with {\em{Spitzer}}, that {\em Herschel} is currently targeting
140 detectable objects, and that only 8 bodies were so far detected at (sub)mm wavelengths.\\ 
\\
As mentioned above, those estimates should not depend on the array configuration, unless the bodies are spatially resolved. In those cases in which the source is spatially resolved, the signal per beam  (lower than the total 
emitted signal) determines the signal-to-noise of a given observation, implying requirements 
higher than 5~$\sigma$ for the thermal detection .
\begin{figure}
\begin{center}
\includegraphics[width=10cm,angle=-90]{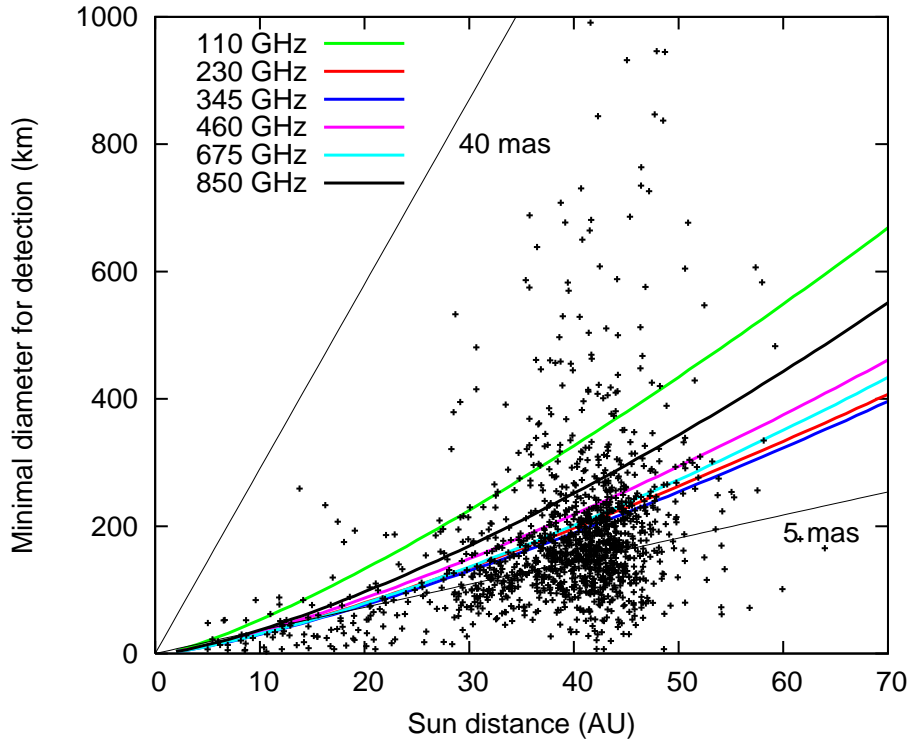} 
\caption{\label{detection} 5~$\sigma$ detection diameter threshold (in km) for detection with ALMA, as a 
function of frequency and heliocentric distance. Black points represent the Centaurs and TNOs 
known as of February 2010 (from Minor Planet Center), along with their measured diameters (when available), or estimated diameters assuming a geometrical albedo of 0.08. Light black lines represent the best spatial resolution available at 110 and 850~GHz (respectively 40 and 5~mas).}
\end{center}
\end{figure}

\subsection{Spatially resolved measurements}
As shown in Figure \ref{detection}, the high angular resolution offered by ALMA at the highest frequencies and most extended configurations (5~mas) allows for the majority of the detectable objects to be spatially resolved, meaning that the synthesized HPBW is smaller or comparable to the object size. When a source is resolved, in addition to total flux measurements, two other methods to analyze the data could be applied: visibility analysis and imaging. The first consists in studying the shape of the visibilities curve in the Fourier plane, where the resolution is apparent when the visibility data vary as a function of projected baseline length (see Figure \ref{charon}), while unresolved sources result in flat visibility functions. The shape of the visibility curve, and in particular the position of the first null, characterizes the brightness temperature distribution and size, and can be compared to synthetic models (e.g. in \citet{moullet2008}). This analysis requires that the HPBW is at least similar to the size of the observed structure. \\
Imaging the data requires a HPBW smaller than the source size, and a satisfying Fourier-plane coverage. On the obtained images, the size, structure and position of the source temperature distribution can be directly observed and measured, with a precision limited by both the synthesized beam and the deconvolution errors. \\
For both visibility analysis and imaging, a SNR {\it{per beam}} of 5 is a minimum, and higher requirements may be put depending on the desired measurement. The spatially resolved projects presented in the following sections require for example a SNR per beam from 6 to 80. Since the signal per beam decreases as the synthesized beam gets smaller with respect to the source, for a given target, the requirements on the rms to be reached are higher than for unresolved observations, implying longer on-source integration time. For example, if the synthesized beam is half the size of the source, one needs an integration time at least 16 times longer for an imaging project than for a spatially unresolved thermal detection at 5~$\sigma$. A compromise between spatial resolution and signal to noise must then be found for each scientific case.
 
\begin{figure}
\begin{center}
\includegraphics[width=10cm,angle=-90]{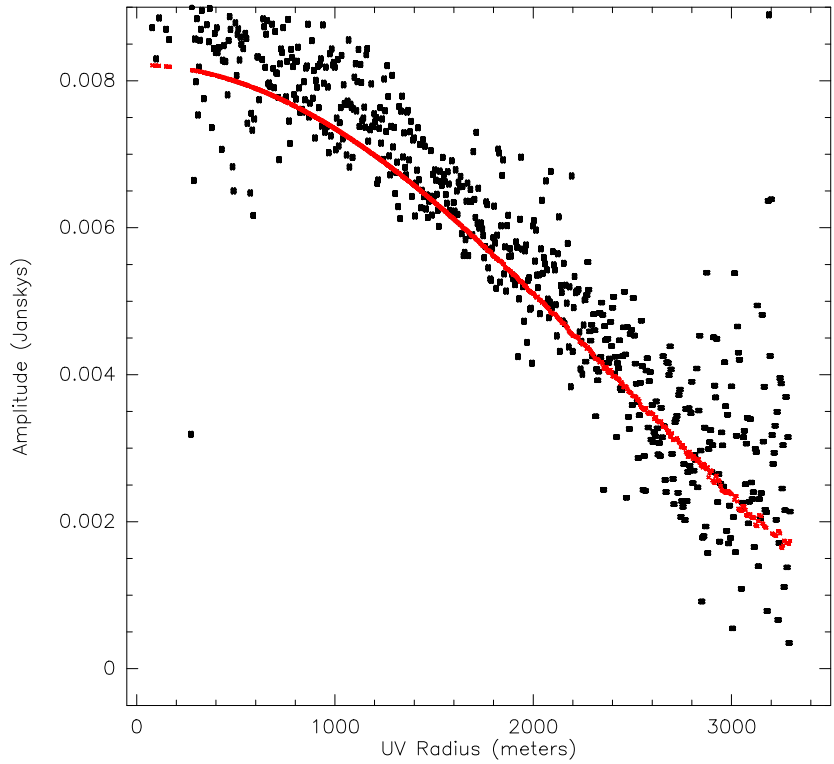}  
\caption{\label{charon} Simulated visibility amplitude curve as a function of projected baseline (dots), for a body of the size (D=1207~km) and distance of Charon, observed during one hour in band 7 (345~GHz) with an extended array configuration. 
The first null is not reached, but the Bessel curve is apparent, as the object is partially resolved. To illustrate the quality of the observations, the expected thermal noise (from Table 1) was applied to the visibilities. The bold line represents the amplitude curve of the model without added noise. }
\end{center}
\end{figure}

\section{Equivalent size and albedo}

Based on the thermal flux estimates and sensitivities presented above, we now describe two different methods to determine
the equivalent size and albedo of an object from ALMA measurements.
We explain hereafter the rationale behind these methods, their assumptions and limitations,
and perform detailed simulations to assess the optimum strategies.

\subsection{Radiometric determination}
As is well known since the IRAS surveys of asteroids \citep{lebofsky1986} and first illustrated by 
\citet{jewitt2001} in the case of TNOs, the combination of optical and thermal flux measurements (radiometric method) permits
in principle to separately determine the albedo and equivalent diameter of airless bodies , i.e. the diameter of a spherical body with the same projected area as the observed object, along with the average brightness temperature. However,
the inferred parameters depend on the details of the temperature distribution models and the uncertainties can be 
prohibitively large (factors of several or even more) if the wavelength of the thermal measurement 
is shorter than the peak of the Planck function \citep{stansberry2008}. These large uncertainties
can be drastically decreased if multi-wavelength thermal infrared measurements are used,
constraining in effect the applicable thermal model (e.g., in an empirical way, through
the beaming factor $\eta$ as defined in Section 3.1), then providing the best accuracy on the temperature and diameter retrieval. Lying in the (quasi)-linear part of the Planck curve, 
the interpretation of single-band (sub)mm-wave 
measurements is not as sensitive to the precise temperature determination as single-band infrared measurements, so that the inferred diameters and albedos are not so much affected by the uncertainty on the thermal model. On the other hand, (sub)mm-wave multi-band measurements will not help to constrain  an appropriate thermal model, but indicate the spectral emissivity.\\
\\
Here we investigate the precision that can be expected on diameter/albedo determination based on ALMA
single-band radiometric thermal flux measurements. Uncertainties result from measurement error and model uncertainty. Measurement
errors are due to the combination of : \\
- the measurement rms noise ($\pm$20\% for a 5~$\sigma$ detection, $\pm$10\% for a 10~$\sigma$ detection): 
this is the uncertainty on the flux emitted by the object, that can be retrieved directly from the visibility data or from the brightness distribution map. Since the choice of the configuration does not have an influence 
on the sensitivity (when the target is not spatially resolved), the measurement of the total
flux can be performed using the most compact configurations, for which phase stability requirements 
are less demanding.\\
- the absolute flux scale error : this could be as low as $\pm$5\% using the calibrations schemes proposed 
by \citet{butler99} or \citet{moreno2002}. \\
The influence of model uncertainty can be assessed by considering the ranges of fluxes predicted for a given
diameter and albedo by the three models described in Section 3.1, spanning a wide range of temperature distributions, and corresponding to surface thermal inertiae from zero to infinite. The effect of the uncertain spectral
emissivity (between 0.8-1) is also included.
With this conservative approach, we obtain flux variations from model to model of about $\pm$25\%,
equivalent to $\pm$12 \% change in diameter and to a $\pm$25 \% change in the albedo as determined
from the visible magnitude. We here neglect any uncertainty on the visible magnitude, 
as the objects in question (i.e. detectable by ALMA) have 
visual magnitudes typically brighter than 21.5, permitting absolute photometry
at the percent level \citep{doressoundiram2005}. To reach this goal, supporting optical observations should be performed on several targets. \\
Combining measurement error and model uncertainty, we find that for the bodies that can be 
detected by ALMA at 5~$\sigma$ or more in the most sensitive bands (6, 7 and 9, i.e. respectively 230, 345 and 675~GHz), the uncertainties on the retrieved 
diameter are between 15-25\%, and are primarily dominated by the model uncertainty ($\sim$14\%). 
The errors reported by recent {\em Herschel}-PACS single-band observations \citep{mueller2010} are of the same order. The total relative error on the diameter obtained after 1 hour integration for bodies at 40~AU is shown in 
Figure \ref{errorbudget}, as a function of diameter. 
 For specific targets, the integration time could be increased to reduce the flux measurements rms, but the result will still be limited by the model uncertainty. \\
The main drawback of the radiometric method then lies in the choice of the thermal model. 
One way to reduce this uncertainty could be the construction of an $ad$ $hoc$ thermal model for each body. The optimal way to do that is to combine ALMA data with {\em Spitzer} photometric data at 24 and 70~microns, and possibly {\em{Herschel}}-SPIRE data. If {\em Spitzer} data is not available, {\em{Herschel}}-SPIRE and PACS photometry, that will be obtained on $\sim$20 objects, can also be used together to constrain the thermal model. The first object to be detected with both instruments, Makemake \citep{lim2010}, revealed that a 2-terrain model is required to fit the thermal fluxes at the observed wavelengths. An alternate approach will be to infer from the combined {\em Spitzer} and {\em Herschel} data, in a statistical sense, the general thermal behavior
of the population (e.g. mean value of $\eta$ and variation trends as a function of orbital parameters) and to use the results for the interpretation of the ALMA data. Finally, for a few objects whose size has been independently derived from a non-radiometric method (imaging, occultation as in \citet{elliot2010}), high-quality thermal measurements can in turn allow to constrain their thermal model and spectral emissivity. 

\begin{figure}
\begin{center}
\includegraphics[width=10cm]{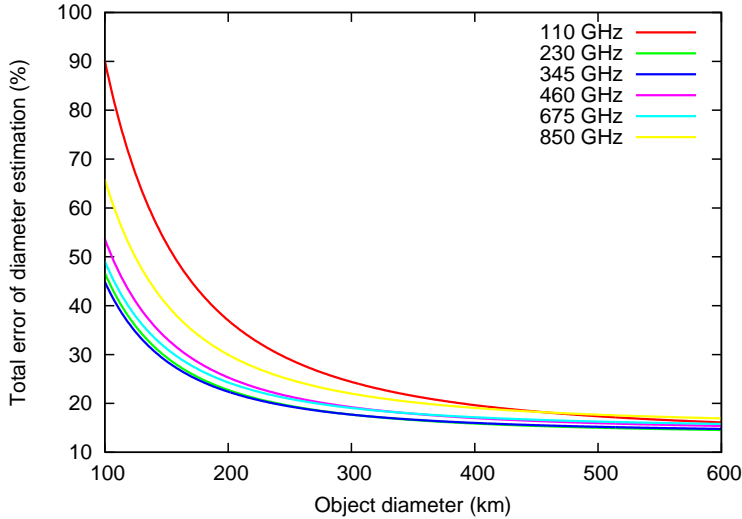}
\caption{\label{errorbudget} Error on the diameter retrieval on targets at 40~AU, as a function of frequency and body diameter, for 1-hour on source. Includes instrumental error and model uncertainty. }
\end{center}
\end{figure}

\subsection{Visibility data analysis}

As mentioned in Section 3.3, the position of the first null of the visibility data curve depends mostly on the size of the emitting region, thus the visibility curve can be analyzed to retrieve this parameter. 
One way to perform this is to fit the visibility data using a disk model : the fitted variable is the equivalent diameter of the source, and the retrieved error bar corresponds to the fit error. Using this direct size measurement, it is then possible to derive the geometric albedo $p_v$ using the visible magnitude.
This technique has already been applied on asteroids Gaspra and Barbara with mid-infrared interferometry 
at the VLTI with a single baseline by \citet{delbo2009}.\\
\\
 To determine which frequency/configuration combination is the most efficient, simulations of observations were performed with the GILDAS ALMA simulator \citep{pety2002}. For an input source model, observing frequency and array configuration for 50 antennas, this tool calculates the expected visibility data with the appropriate Fourier-plane coverage and noise level. Here all simulations assumed a spherical target at -10$^{\circ}$ declination, with a uniform temperature distribution and no emissivity variations (Lambertian surface model). Fitting of the simulated visibility data with a plain disk model was performed through a $\chi^2$ fit automated in GILDAS. The error bars were estimated on the one hand from the GILDAS routine and on the other hand from the difference
between the known (input) diameters and those retrieved from the simulator.\\
A series of simulations shows that the choice of an appropriate configuration and frequency clearly depends on the target size and brightness temperature, as the quality of the fit is driven by both SNR and spatial resolution. The best way to determine the observing strategy for a given project is then to perform such simulations for each target. We could notice that a minimal SNR per beam of $\sim$20 is necessary to perform the fit. In addition, the best results on diameter determination are constantly obtained when the synthesized beam size resolution is comparable to the size of the target (between 0.6-1.2 times the size of the target), corresponding to an important concentration of visibilities in a region of the Fourier plane that is particularly sensitive to the body's size. For example at 45~AU, for targets larger than 26~mas, it appears that the best results are 
achieved with band 7 (345~GHz) with very extended configurations (baselines $>$ 6000~m). Good results are also obtained with bands 9 and 10 (675 and 850~GHz), for which baselines of 
3000~m can be sufficient. Those however require significantly better weather conditions.  For targets with lower sizes, the highest frequencies (bands 9 and 10) provide 
the best results, combined with array configurations of $\sim$3000 to 6000~m
 extent. We note that, although observations at the highest frequencies with the most extended configurations 
permit the highest spatial resolution, they do not necessarily provide the best results, either because they resolve the source too much or because they do not provide the best SNR per beam.\\
\\
 By observing with an appropriate frequency and configuration, this technique allows the retrieval of sizes with an accuracy better than or comparable to that achieved from the total flux radiometric technique for a number of targets, whose apparent size can be as low as 15~mas.
The accuracies found on the retrieved diameters are of the order of 1.5\%  for large (1100-1500~km) and thermally bright bodies such as Quaoar, Makemake, Haumea and 2002TC$_{302}$, when observed with band 7. Errors lower than 7\% should be achieved on 2002AW$_{197}$, Orcus, Varuna, Chariklo, 2004GV$_9$, Huya and 2003VS$_2$ with observations in bands 7, 9, or 10, and uncertainties lower than 15\% are expected for medium-sized (700-200~km class) or very cold bodies, after at least 2 hours of integration in band 10. An improved determination of the equivalent diameters, and hence albedos, is then possible on at least 30 bodies, whose diameters are presently known with a 10-30\% uncertainty. Compared to the
radiometric technique presented in Section 4.1, results are not as much affected by the uncertainty on the thermal model, so that the diameters retrieved can in turn be used to interpret the simultaneously obtained thermal flux in terms of thermal model and emissivity.  However the  size of the thermal emission is not necessarily identical to the size of the solid body, since the brightness temperature distribution is not uniform over the disk. The difference is in fact again 
model-dependent, and is estimated to account for an additional 4\% model uncertainty that must be added to the fit 
error. This difference is mostly important at the highest frequencies, since lower frequencies are not as sensitive to the surface temperature distribution.  For bodies whose sizes are already known to very good 
accuracy (e.g. Charon from stellar occultations, see \citet{sicardy2006}), the measurement of the 
characteristic thermal emission size can be an excellent diagnostic of brightness temperature
distribution, especially diurnal, latitudinal and with emission angle, that can be in turn tied to properties such 
as thermal inertia and surface dielectric constant. Finally, this technique cannot be applied directly to multiple systems, for which the separation between members must be known to analyze the visibility curve. This parameter can be estimated by mapping the system, as presented in Section 7. \\

\section{Shapes}
Both methods presented in Section 4 assume spherical shapes and hence can only determine the 
 equivalent diameter of a body. Although this can be a good approximation for the largest, slowly
rotating bodies, many objects are suspected to have elongated or irregular shapes. Since bodies larger than $\sim$50~km are thought to be controlled by hydrostatic equilibrium, their shape 
is expected to be close to a Jacobi ellipsoid, and determination of their axes ratio can give clues on the material strength and bulk density \citep{farinella1987,lacerda2007}. For smaller bodies, irregular shapes
can  be the result of collisions. \\
Shapes are in general inferred from the observed amplitude and period of optical rotational lightcurves  
\citep{sheppard2008}. Lightcurves result from the variation of the apparent projected area 
with rotational phase, giving access to the projected axes ratio, corresponding to a lower limit on the ratio of the two axes perpendicular to the rotation axis of the body.  The third (polar) axis is determined from hydrostatic
equilibrium.  However, albedo markings may affect the shape of lightcurves
(see \citet{lacerda2008} for Haumea) or entirely dominate it (Pluto, \citet{buie1997}). One way to disentangle 
between shape and albedo effects in visible lightcurves is to observe thermal lightcurves, which
depend on disk-averaged albedo variations in an anti-correlated and, except for very reflective objects, less sensitive manner than optical lightcurves (see e.g. the absence of detected lightcurve amplitude at 1.2~mm on Pluto by \citep{lellouch2000b}), since the surface temperature varies as $(1-A_b)^{0.25}$, and $A_b$ is often small ($\ll$1). \\
\\
With the sensitivity of ALMA it is possible to sample thermal 
rotational lightcurves on a large number of Centaurs and TNOs, whose rotation periods are on  average 
$\sim$8~hours \citep{sheppard2008}, though not with a time resolution as high as in the
optical. As for optical lightcurves, the amplitude of the thermal lightcurve is mostly related to the projected axes ratio.
 Note however that for a given shape, the amplitude of the thermal lightcurve is usually
larger than that of the optical lightcurve. This stems from the fact that the distribution of solar zenith
angles and hence temperatures over the visible disk are much "flatter" at lightcurve maximum than at the 
minimum. A recent example is given by the {\em Herschel}-PACS
observations of Haumea \citep{lellouch2010}.  The effect is however subdued at the longest wavelengths
and we ignore it in what follows. 
\\
For all targets that can be detected at 5~$\sigma$ with band 7 (345~GHz), variations of flux larger than 40\% can be assessed, 
corresponding to a projected axes ratio ${a}/{b}>1.4$. For more than 200 large bodies, detectable 
at a 10~$\sigma$ level in less than 1~hour, projected axes ratio larger than 1.2 could be identified. Comparing these results with the list of bodies with a known rotational lightcurve in the optical/near-infrared \citep{sheppard2008}, we can see that 
significant variations on thermal lightcurves could be observed with ALMA on at least 30 bodies,  then allowing to put tighter constraints on the presence of albedo markings and on their projected axes ratio.\\
The lightcurve method is however limited to bodies presenting a favorable geometric configuration, and
in particular cannot detect ellipticity for pole-on orientation or for bodies with long (or very short) rotation
periods. \\
\\
The alternative method to determine shapes is to identify ellipticity directly on high spatial resolution interferometric data recorded at one given
rotational phase, through 
visibility data fitting with an elliptic model. The inferred parameters in this case are the  
semi-major and semi-minor axes in the sky plane and the position angle of the ellipse at the relevant phase, along with 
their error bars. Typically, fitting errors are approximately a factor $\sqrt{2}$ larger than the fit errors for a simple disk model fit (see Section 4.2). Considering that
the elliptical shape can be established if the retrieved axes differ within the error bars, 
we find that projected axes ratios down to 1.04 for the 6 largest bodies (in two hours from band 7 observations, i.e. 345~GHz), and down to 1.2 for $\sim$7 additional bodies (from observations in bands 10, 9, or 7, i.e. 850, 675, or 345~GHz), can be detected. The method is therefore applicable
to fewer objects than the thermal lightcurve approach. However it appears complementary as the 
measured axes ratios are projected in a different plane. For example, for an equator-on configuration,
the thermal lightcurve method determines the two equatorial axes, while the "direct imaging" method
determines the polar axis and a combination of the two equatorial axes that depends on the rotational
phase. In addition, this is the only applicable method to measure non-sphericity for large objects without any measured 
lightcurve, e.g. Makemake, Orcus and 2002 AW$_{197}$, whose potential ellipticity may not have been detected because of an unfavorable geometry (pole-on inclination of the rotation axis).

\section{Surface mapping}
From infrared and visible spectroscopy, the surfaces of Centaurs and TNOs have revealed a variety of surface materials. Some bodies display spectral bands characteristic of ices (mostly H$_2$O and CH$_4$), while others are featureless, with broadband colors suggesting carbon and organic rich materials \citep{barucci2008}. 
A handful of these objects have also revealed rotational variations of their spectra, suggesting an heterogeneous 
surface. This is in particular the case of Pluto, for which, based on the variability and monitoring
of the visible and near-infrared spectrum, \citet{grundy1996} and \citet{grundy2001} proposed
a possible distribution of tholins, water and methane ices, in general agreement with
direct HST imaging \citep{stern1995,buie2010a} and optical lightcurves \citep{buie1997,buie2010b}.
Possible causes behind these surface variations include the resurfacing of fresh and bright ice due to collisions, cryo-volcanism and atmospheric condensation/sublimation
cycles, while surfaces exposed for a longer time get redder and darker due to space weathering and irradiation.
So far such variations could not be assessed on other bodies, due to the lack of spatial resolution and the
paucity of rotationally resolved spectra. \\
Brightness temperature maps of Centaurs or TNOs have to date never been obtained, and could help to distinguish surface features. Although the thermal emission of those bodies is typically only weakly dependent on their albedos, this is not true for the most reflective of them. For those bodies,
maps of the thermal emission should permit to map surface reflectivity. In addition, even if small-scale
variations of the brightness temperature cannot be distinguished (e.g. generally  due to too dark a surface), the overall limb-to-limb brightness
temperature distribution (e.g. diurnal and equator-to-pole) will be diagnostic of the thermal
inertia. In the case of Pluto, thermal models predict, based on ISO observations, very high variations of the surface temperature, up to 50\%, produced by the combination of high albedo variations and response to insolation taking into account the thermal inertia \citep{lellouch2000}. We will present here the case of maps where the surface is resolved in at least two points in each direction. \\
Thermal flux variations of 10\%, corresponding to brightness temperature variations of 5-10\%, could be detected with a SNR per beam of 30. With this SNR requirement, mapping is possible on 6 large objects  with a 4-hour long observation in bands 7, 9 and 10 (345, 675 and 850~GHz), with a spatial resolution rom 17 to 14~mas. To illustrate the effect of Pluto's temperature variations in ALMA
observations, we used the thermal map of \citet{lellouch2000}, calculated from the surface model 
of \cite{grundy1996} at an orbital longitude of 93$^{\circ}$W.  Figure \ref{pluto} shows synthesized maps 
 obtained using band 7 and band 10 with a very extended configuration. Such observations would provide the first direct thermal mapping of Pluto.\\
\\
Search for spatial albedo variations based on brightness temperature mapping is difficult because they produce relatively small variations of the thermal emission. Indeed, assuming the average phase integral value of 0.4, even large albedo variations (from p$_v$=0.7 to p$_v$=1) can only produce 5\% variations on the brightness temperature, or, assuming a phase integral of 0.8 (corresponding to the value estimated for Pluto by  \citet{lellouch2000}), 20\% at most. Such brightness temperature variations would result at most in variations of 5-40\% of the thermal emission, depending on the observing frequency.\\
 We consider here that only variations higher than 3~$\sigma$ on the thermal emission can be detected, and we assume that the albedo is the only property of the surface that could be varying. If q=0.4, a SNR as a high as 80 is necessary to retrieve albedos variations greater than 15\% (band 10) or 20\% (band 7). Among the objects that can be mapped, such a SNR per beam can only be obtained, in four hours of observation, on a few very large bodies, i.e. Charon, Quaoar and Makemake. The spatial resolution of 28~mas (band 10), 22~mas (band 9) or 20~mas (band 7), is limited by the SNR per beam, and allows a 2x2 mapping only. Such observations in bands 9 and 10 should be performed with at least 3-4~km wide configurations, while band 7 observations must make use of with very extended ($\sim$8~km) configurations, although with lower sky opacity requirements than higher frequencies. If q=0.8, a SNR of 30 can allow to retrieve albedo variations of 12\% (band 10) to 15\% (band 7). This SNR can be obtained on 6 large bodies, with a spatial resolution down to $\sim$ 15~mas. Although no fine albedo mapping with this method is possible, it could reveal a strongly inhomogeneous albedo distribution on Charon and Makemake, as was found on Pluto.\\ We note that in July 2015, during the New Horizons encounter \citep{young2008}, the 
LEISA near-infrared spectrometer, part of the Ralph instrument suite \citep{reuters2008}, will use temperature-sensitive N$_2$, CH$_4$ and H$_2$O bands to map surface temperatures, while the radiometric mode of the Radio experiment package (REX) will be used to obtain a direct measurement of Pluto's surface  temperature. Although the New Horizons observations will obviously
afford much higher spatial resolution, the ALMA maps will be complementary, noting in particular that
LEISA may not be able to determine temperature of the possible ice-free regions.
\\

\begin{figure}
\begin{minipage}{18cm}
\includegraphics[width=7cm,angle=-90]{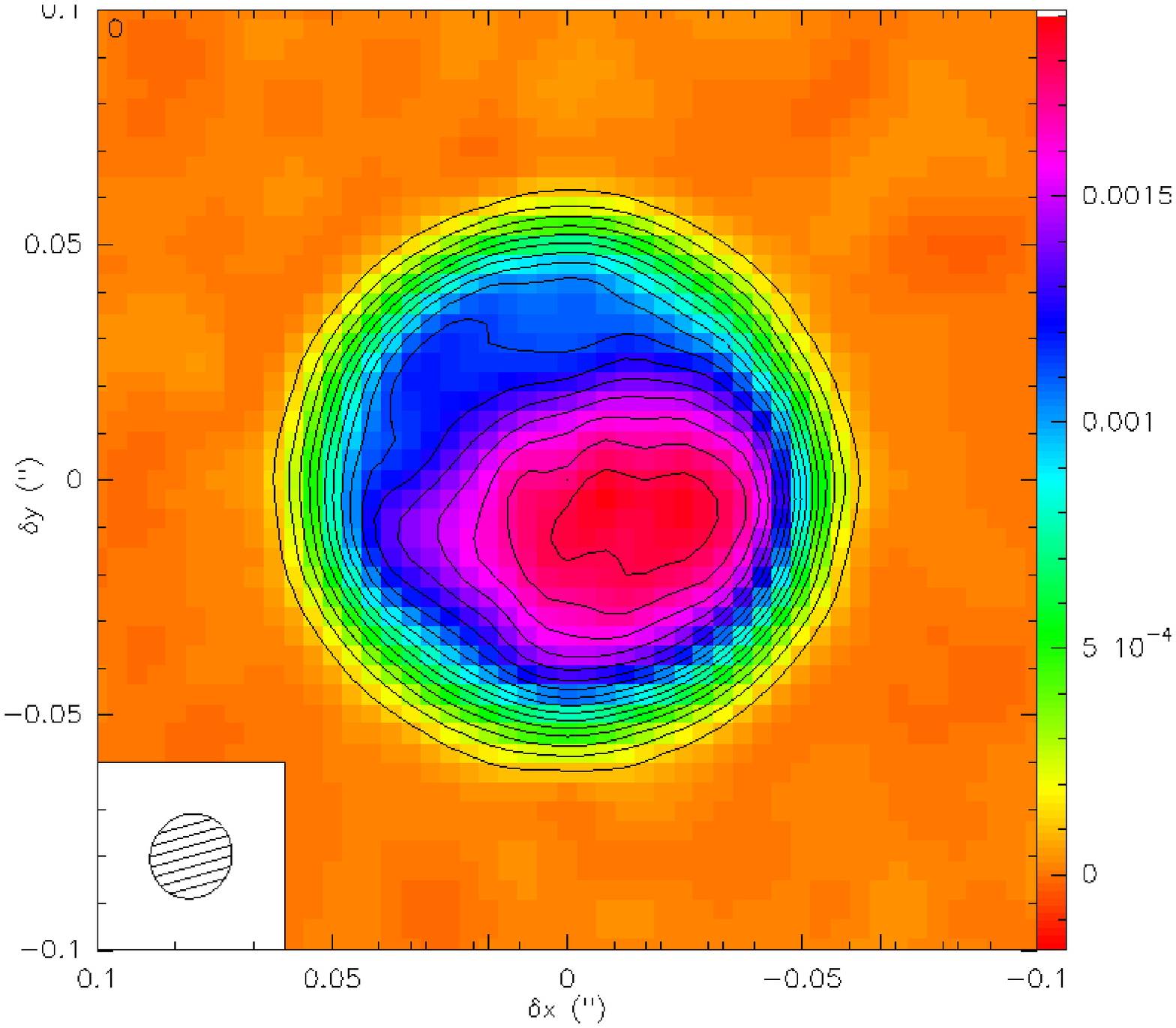}
\includegraphics[width=7cm,angle=-90]{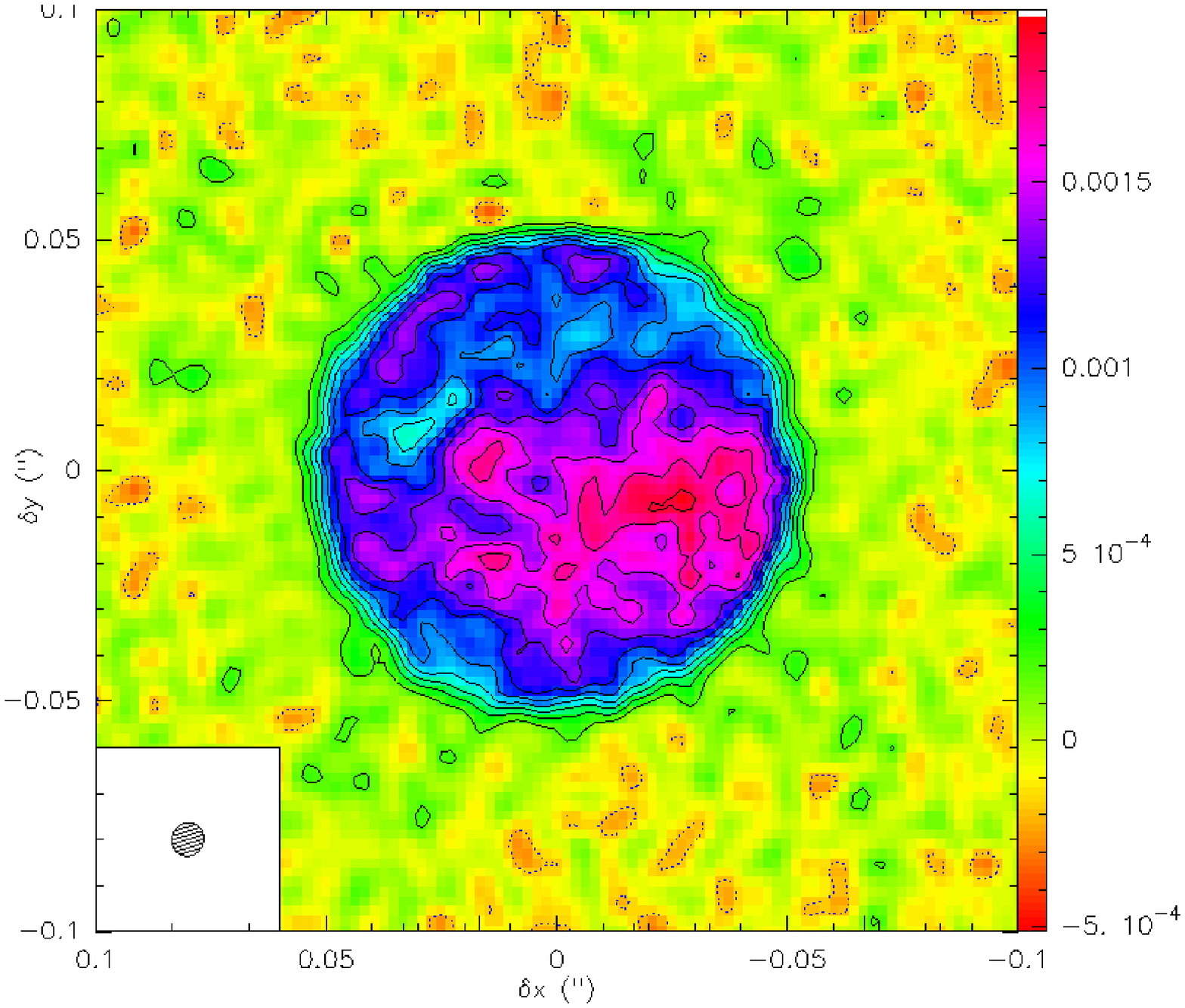}
\end{minipage}
\caption{\label{pluto} Synthetic maps of a simulated observation of Pluto, assuming the thermal model from 
\citet{lellouch2000} based on the \citet{grundy1996} albedo map, for a 4-hour observation with a very extended 
(10~km) configuration. Left : observations at 345~GHz. Contours are set to 10~$\sigma$, the beam size is 24~mas. Right : observations at 850~GHz. Contours are set to 2~$\sigma$, the beam size is 10~mas. The synthesized beam is represented in the inset in the bottom-left corner. Axes are expressed in arcseconds. The color scale (Jy/beam) is indicated by the vertical bar on the right. }
\end{figure}

\section{Multiple system imaging}
\begin{figure}
\begin{minipage}{18cm}
\includegraphics[width=7cm,angle=-90]{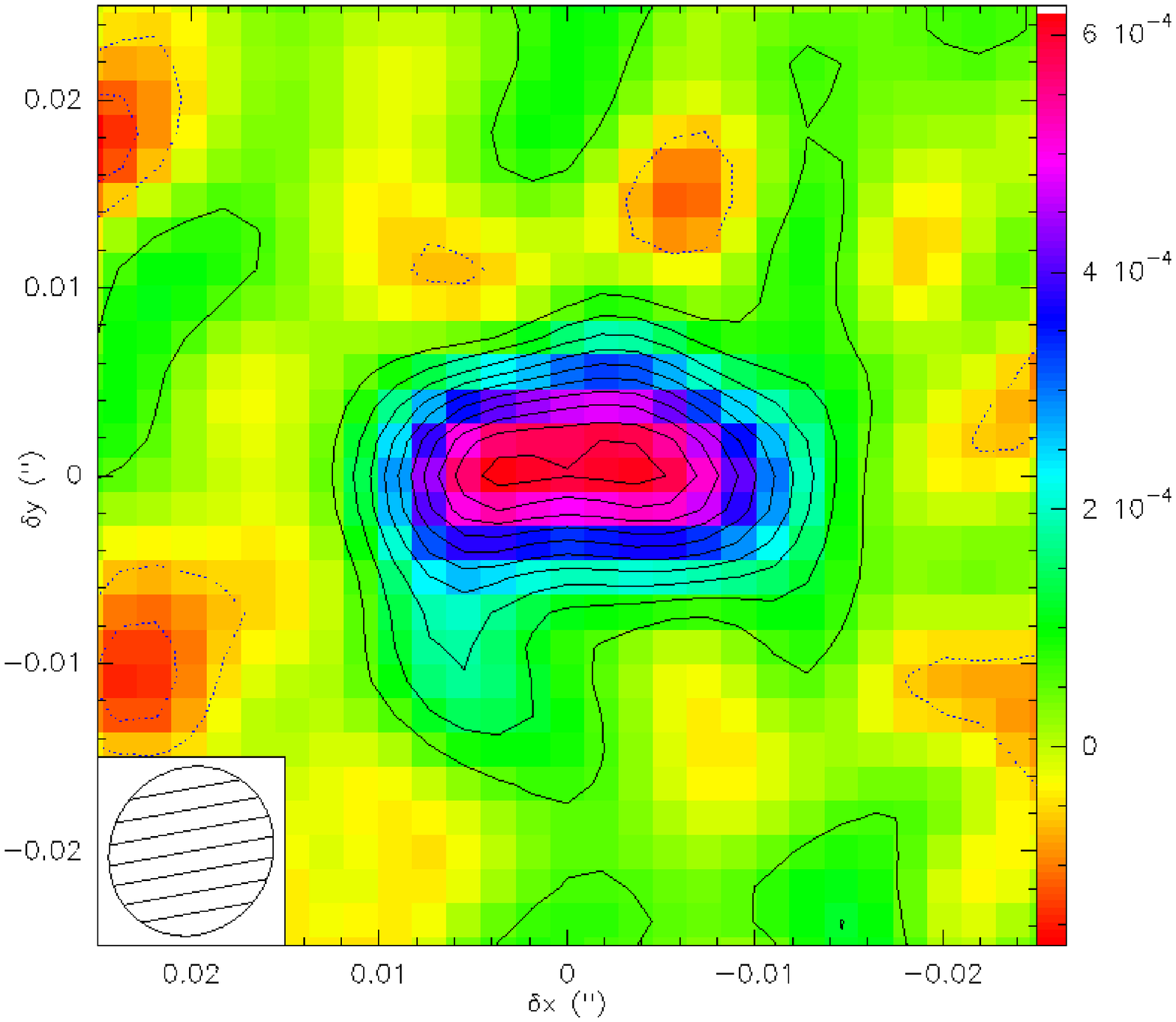}
\includegraphics[width=7cm,angle=-90]{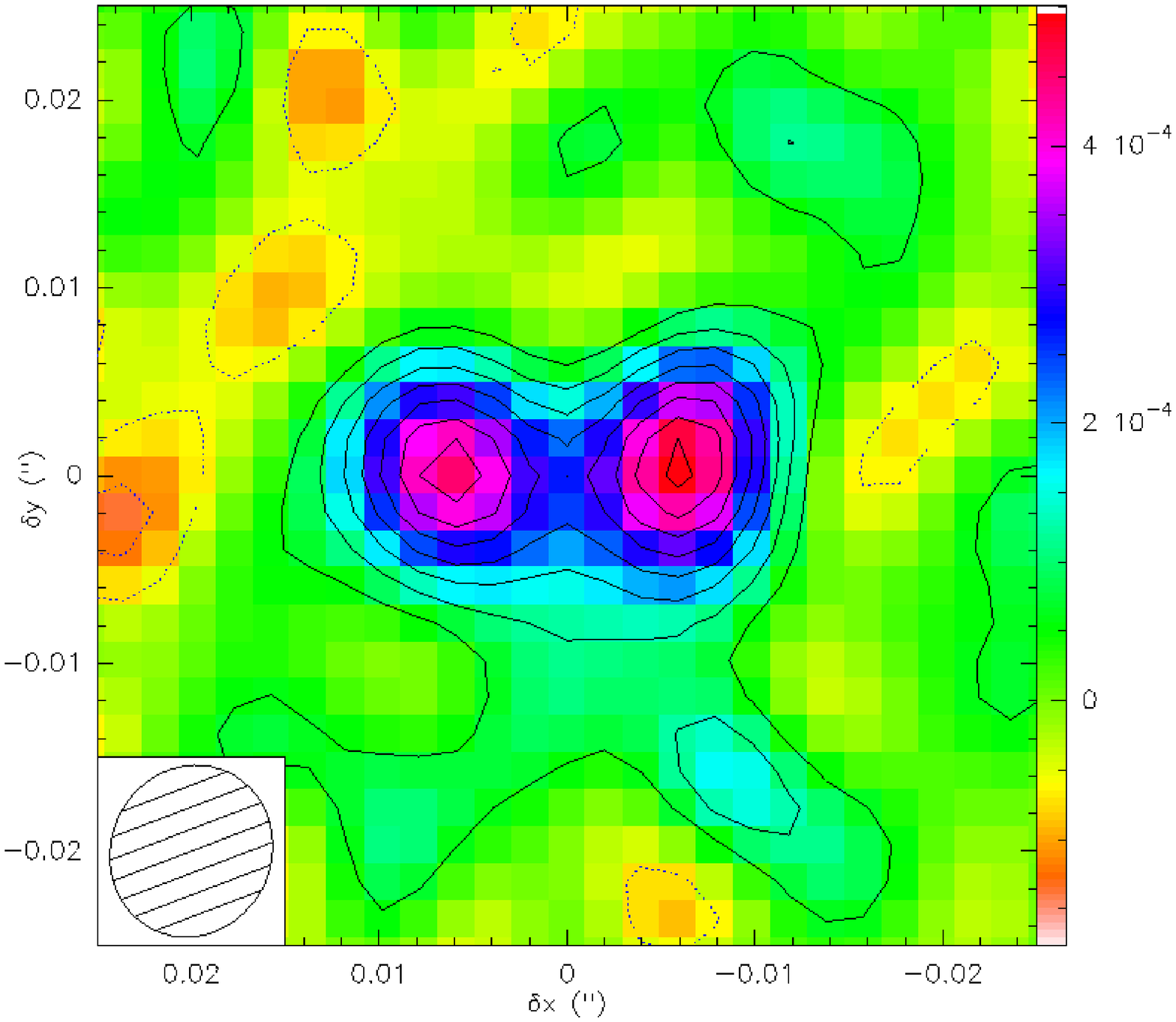}
\end{minipage}
\caption{\label{binary} Synthetic maps of a simulated observation of a binary system at 675~GHz with a 10~km wide configuration. Left : two bodies of 8~mas apparent size at 30~AU (174~km diameter), separated by 10~mas. Right : the same bodies with a separation of 12~mas. Contours are set to the rms reached after 2 hours on source (6$\times$10$^{-5}$ Jy/beam). The synthesized beam (12~mas) is represented in the inset at the bottom-left corner. Axes are expressed in arcseconds. The color scale (Jy/beam) is indicated by the vertical bar on the right.\label{lastfig} }
\end{figure}
Large  imaging surveys, mostly with HST, have revealed that more than 50 TNOs and Centaurs are actually multiple systems, representing a rather high fraction of the whole population of more than 1400 objects \citep{noll2008}. Most of those consist in nearly equally-sized binaries that suggest formation by dynamic capture, but the measured size ratios can go as low as 1:100. Studying these systems, while leading to insights into the processes of capture and fragmentation, also provides
 the only access to the system's total mass through mutual orbit determination. If the individual sizes
(or at least the total equivalent size) of the system members are known, this leads to determination of the average density, bringing a strong constraint on interior composition and structure. Bulk densities could also be correlated to other properties such as size, albedo and heliocentric distance.\\
\\
In spite of obvious observational biases, which  favor systems with high separations,
the frequency of binaries seems to increase as the separation decreases \citep{noll2008}, warranting the use
of the highest possible spatial resolution.
At best, the point spread function of HST/STIS allows the detection of equally bright pairs separated by 30~mas, and to determine the separation with an error bar larger than 2~mas.\\
Although one may not immediately think of (sub)mm-wave observations as the most obvious method 
to enhance spatial resolution, the high angular resolution of ALMA (see Figure \ref{resolution}) makes it 
competitive with the most performing space telescopes on this particular topic. To estimate the 
separation power of ALMA for binary Centaurs and TNOs, simulations were performed for two-hour observations of 
different sets of equally-sized binary systems. We adopted the reference thermal model of Section 3.1
and an albedo of 0.08. We considered that a binary system is resolved when the emission map
shows two emission peaks with significant separation (i.e. with a minimum emission level 
at least 3~$\sigma$ lower than the peaks between them), as is illustrated in Figure \ref{binary}. This conservative requirement could be relaxed if the image fidelity, i.e. the imaging quality, is good enough so that the errors due to image deconvolution are low.\\
\\
We observe that, provided that each member of the system is detected with a SNR ratio higher than 6, the 
separation power of the array is equal to its synthesized beam size.
As an example, for a configuration with a maximum 10~km baseline, the separation power reaches 10~mas in band 10 (850~GHz), 12~mas in band 9 (675~GHz), 18~mas in band 8 (460~GHz), and 24~mas in band 7 (345~GHz). 
However this does not mean that choosing the highest frequency is always the best strategy. Indeed 
observations in band 7 are more sensitive than observations in band 10, giving access to smaller targets. 
In band 10 and for a 30~AU distance, in two hours on source, only (individual) targets with apparent sizes larger than 9.5~mas have a sufficient thermal flux, given our assumed thermal model, to be detected at 6~$\sigma$. Band 10 could then be possibly used only for a selection of rather large targets whose multiplicity 
is either highly suspected or already assessed (e.g. 1997 CS$_{29}$, 2003 QA$_{91}$, 2001 QY$_{297}$), 
so as to get a very good accuracy on their separation. For band 9 and band 7 observations, at 30~AU, targets larger than respectively 6 and 5.5~mas are suitable, 
and with this limitation, assuming that all known bodies are equally-sized and equally-bright binary systems, 
as many as 250 among them would be bright enough for each member to be detected and thus binarity to
 be determined. \\
Once the multiplicity of a system is assessed, it is possible to retrieve its projected separation by measuring 
directly on the map the relative distance between the two peaks. Following astrometric error budgets from 
\citet{lestrade2008}, we can estimate that with a minimal SNR of 6 on peak, 
the astrometric error on each peak position will be lower than 0.7~mas at 345~GHz, and 
 0.3~mas at 850~GHz, the latter being comparable to the astrometric precision of the FGS on HST. Including deconvolution effects, we estimate conservatively 
that the error on the separation can be lower than 1.5~mas, corresponding to $\sim$40~km at 40~AU. \\
\\
Given that the configuration of the system at the moment of detection is unknown, 
the obtained projected separation is only a lower limit to the actual system separation. 
Only follow-up observations can allow the retrieval of the separation and the orbiting period 
\citep{hestroffer2005}, that are necessary to derive the total system mass. 
In addition, independent measurements of the flux density of each body can be obtained. If the system is resolved both at visible
and thermal wavelengths, it is possible to retrieve each member's 
diameter using the radiometric method described in Section 4.1. If no visible magnitude for each member is available, 
an acceptable estimate of the individual sizes can be obtained assuming the albedo value, or using the whole system 
magnitude to derive an average albedo value.\\

\section{Conclusions}
The calculations and simulations presented in this paper show that the capabilities of 
ALMA at completion in terms of sensitivity, spatial resolution and imaging will be well 
suited to study a large number of Centaurs and TNOs.
\\
Determination of equivalent diameters and albedos via the radiometric method may target more than 500 objects using spatially unresolved continuum measurements in bands 6 or 7, allowing improved determination of the size distribution for bodies larger than 100~km, and helping to provide robust conclusions when correlations between physical, spectral and dynamical properties are examined.\\
With the same bands, thermal rotational lightcurves could be sampled on $\sim$30 targets that present a significant optical lightcurve, allowing one to disentangle the effects of non-spherical shape and surface albedo variations. \\
Using extended configurations, 
bands 7, 9 and 10 could provide essentially model-independent size and ellipticity measurements, 
using visibility data fitting, on $\sim$30 large bodies. The first thermal maps of the 6 largest objects could also be obtained with a spatial resolution down to $\sim$14~mas.\\
Finally, ALMA will be the only facility that could separate multiple systems as close as 10~mas, and possibly contact binaries. Retrieval of the system orbit and period is key for the determination of the system's mass. \\
\\   
Although the first interferometric fringes with a 3-antenna array on site were obtained in 
November 2009, the completion of ALMA is a few years away. 
A first call for proposal on early science will be issued in spring 2011 for a partial array ($\sim$16 antennas), that will be sufficient for the detection of the largest sources. 

\section*{Acknowledgments}
We thank the two anonymous referees for their thorough review of the paper. We also thank S. Guilloteau and A. Dutrey for their useful comments and tips, M. Holman for his proofreading and interesting discussions, J. Berthier from IMCCE for his help on ephemeris computation and the GILDAS team for their support.

\label{lastpage}
\bibliographystyle{biblioicarus}
\def\aj{AJ}   
\def\aap{A\&A}   
 \def\apjl{ApJ}   
 \def\nat{Nature}              % Nature       
\def\apj{ApJ} 

\bibliography{bibliography.bib}

\end{document}